\documentclass[a4paper,6pt,intlimits]{iopart}
\expandafter\let\csname equation*\endcsname\relax
\expandafter\let\csname endequation*\endcsname\relax
\usepackage{graphicx}
\usepackage{amsmath}           
\usepackage{amsfonts}          
\usepackage{amssymb}           
\usepackage{epstopdf}

\usepackage{tikz,bbm,bm,mathrsfs,dsfont}
\usetikzlibrary{shapes,arrows,pgfplots.groupplots,fadings,calc,decorations.markings,positioning,chains,fit,shapes,calc}

\usepackage{color}
\usepackage{quoting}

\usepackage{hyperref}
\hypersetup{
 colorlinks=true}
\usepackage{cite}

\usepackage{dsfont}

\newcommand{\be}{\begin{equation}}
\newcommand{\ee}{\end{equation}}
\def\reff#1{(\protect\ref{#1})}

\numberwithin{equation}{section}

\begin{document}
\title{Selberg integrals in 1D random Euclidean optimization problems}
\author{Sergio Caracciolo}\ead{sergio.caracciolo@mi.infn.it}
\address{Dipartimento di Fisica, University of Milan and INFN, via Celoria 16, 20133 Milan, Italy}
\author{Andrea Di Gioacchino}\ead{andrea.digioacchino@unimi.it}
\address{Dipartimento di Fisica, University of Milan and INFN, via Celoria 16, 20133 Milan, Italy}
\author{Enrico M. Malatesta}\ead{enrico.m.malatesta@gmail.com}
\address{Dipartimento di Fisica, University of Milan and INFN, via Celoria 16, 20133 Milan, Italy}
\author{Luca G. Molinari}\ead{luca.molinari@mi.infn.it}
\address{Dipartimento di Fisica, University of Milan and INFN, via Celoria 16, 20133 Milan, Italy}
\date{\today}
\begin{abstract}
We consider a set of Euclidean optimization problems in one dimension, where the cost function associated to the couple of points $x$ and $y$ is the Euclidean distance between them to an arbitrary power $p\ge1$, and the points are chosen at random with uniform measure.
We derive the exact average cost for the random assignment problem, for any number of points, by using Selberg's integrals. 
Some variants of these integrals allow to derive also the exact average cost for the bipartite travelling salesman problem.
\end{abstract}


\section{Selberg integrals}

Euler Beta integrals
\be
\hbox{Beta\,}(\alpha, \beta):=  \int_0^1 \ dx \, x^{\alpha-1} (1-x)^{\beta -1} = \frac{\Gamma( \alpha) \Gamma(\beta)}{\Gamma(\alpha+ \beta)} \label{beta}
\ee
with $\alpha, \beta \in \mathbb{C}$ and $\Re(\alpha)>0$, $\Re(\beta)>0$, have been generalized in the 1940s by Atle Selberg~\cite{Selberg}
\begin{equation}
\begin{split}
S_n(\alpha, \beta, \gamma) & :=  \left(\prod_{i=1}^n \int_0^1 \ dx_i \, x_i^{\alpha-1} (1-x_i)^{\beta -1}\right) \left |\Delta(x)\right |^{2 \gamma} \\
& = \prod_{j=1}^n \frac{\Gamma( \alpha + (j-1) \gamma) \Gamma(\beta + (j-1) \gamma)\Gamma(1 + j \gamma)}{\Gamma(\alpha+ \beta + (n+j-2) \gamma)\Gamma(1 +  \gamma)} \label{S_n}
\end{split}
\end{equation}
where
\be
\Delta(x) := \prod_{1\leq i < j \leq n} (x_i-x_j)  \, 
\ee
with $\alpha, \beta, \gamma \in \mathbb{C}$ and $\Re(\alpha)>0$, $\Re(\beta)>0$, $\Re(\gamma) > \min(1/n, \Re(\alpha)/(n-1), \Re(\beta)/(n-1) )$,
see~\cite[Chap. 8]{Special}. Indeed~\reff{S_n} reduces to~\reff{beta} when $n=1$.

These integrals have been used by Enrico Bombieri (see~\cite{Forrester} for the detailed history) to prove what was known as the {\em Mehta-Dyson conjecture}~\cite{Dyson, Mehta67, Mehta74, Mehta04} in random matrix theory, that is
\be
F_n(\gamma) :=   \left(\prod_{i=1}^n \int_{-\infty}^{+\infty} \, \frac{dx_i}{\sqrt{2 \pi}} \, e^{-\frac{x_i^2}{2}} \right)\left |\Delta(x)\right |^{2 \gamma} 
=  \prod_{j=1}^n \frac{\Gamma(1 + j \gamma)}{\Gamma(1 +  \gamma)} \label{F_n} \, 
\ee
but they found many applications also in the context of conformal field theories~\cite{Dotsenko} and exactly solvable models, for example in the evaluation of the norm of {\em Jack polynomials}\cite{Kakei}.
In the following we shall also need of an extension of Selberg integrals~\cite[Sec. 8.3]{Special}
\begin{equation}
\begin{split}
B_n(j, k; \alpha, \beta, \gamma) & :=  \left(\prod_{i=1}^n \int_0^1 \ dx_i \, x_i^{\alpha-1} (1-x_i)^{\beta -1}\right)\left( \prod_{s=1}^j x_s\right) \left( \prod_{s=j+1}^{j+k} (1-x_s)\right) \left |\Delta(x)\right |^{2 \gamma} \label{B_{j,k}}\\
& = 
\, S_n(\alpha, \beta, \gamma)\, \frac{\prod_{i=1}^j [\alpha +(n-i)\gamma] \prod_{i=1}^k [\beta +(n-i)\gamma] }{\prod_{i=1}^{j+k} [\alpha+\beta +(2n-1-i)\gamma]} \, .
\end{split}
\end{equation}

In this paper we present one more application of Selberg integrals in the context of Euclidean random combinatorial optimization problems in one dimension. 
While these problems have been well understood in the case in which the entries of the cost matrix, that is the costs between couple of points,  are independent and equally distributed random variables, also thanks to the methods developed in the context of statistical mechanics of disordered systems~\cite{mezard1987spin}, the interesting physical case in which the cost matrix is a Euclidean random matrix~\cite{MPZ} is much less studied because of the difficulties induced by correlations among the entries.
The possibility of dealing with correlations as a perturbation has been investigated~\cite{Mezard1988} and is effective for very large dimension of the Euclidean space. On the opposite side, detailed results, at least in the limit of an asymptotic number of points, have been recently obtained in one dimension~\cite{Caracciolo:159, Caracciolo:160, Caracciolo:169, Caracciolo:171, Caracciolo:173,Caracciolo2018} and in two dimensions~\cite{Caracciolo:158, Caracciolo:162, Caracciolo:174, Ambrosio2016, CaraccioloSicuro2015}. Note that related problems in higher than one dimension have also been studied from a rigorous point of view~\cite{Talagrand1994,Trillos2015,Trillos2016}.

We shall concentrate here in the one dimensional case, with the aim at proving new exact results, which we have obtained by exploiting the Selberg integrals.
Let us consider $N$ points chosen at random in the interval $[0,1]$. Let us order them in such a way that $x_i < x_{i+1}$ for $i=1, \dots, N-1$. The probability of finding the $k$-th  point in the interval $[x, x+dx]$ is
\be
P_k(x)dx := \frac{\Gamma(N+1)}{\Gamma(k) \, \Gamma(N-k+1)} \, x^{k-1} (1-x)^{N-k} dx \,.
\label{P_k}
\ee
To every couple $x_i$ and $x_j$ we associate a cost which depends only on their Euclidean distance, in the form
\be
w_{ij}^{(p)} = f(| x_i - x_j |) := | x_i - x_j |^p
\ee
for some $p > 1$, so that the function $f$ is an increasing and convex function of its argument.

The new results presented in this paper regard finite-size systems. The paper is organized as follows. In Sec.~\ref{2}, respectively Sec.~\ref{3}, we present the evaluation of the average cost for the assignment, respectively the TSP problem, for all values of the parameter $p>1$ and all numbers of points. In Sec.~\ref{4} we evaluate, instead, under the same conditions, what is lost in the average cost if we do not match the $k$-th  point in the best way, for each $k$.  This is needed if we want to compute upper bounds to the average cost of the 2-factor problem.

\section{Average cost in the assignment }\label{2}

Consider two sets of $N$ points chosen at random in the interval $[0,1]$, the red $\{r_i\}_{\in[N]}$, and the blue $\{b_i\}_{i\in[N]}$, points. In the assignment problem we have to find the one-to-one correspondence between the red and blue points, that is a permutation $\sigma$ in the symmetric group of $N$ elements ${\mathcal S}_N$, such that the total cost
\be
E_N^{(p)}[\sigma] := \sum_{i=1}^{N} w_{i,\sigma(i)}^{(p)} = \sum_{i=1}^{N} | r_i - b_{\sigma(i)} |^p
\ee
is minimized. 

When $p>1$ the optimal solution is the identity permutation, once both sets of points have been ordered~\cite{McCannRobert1999, Caracciolo:159}. It follows that, in these cases, the optimal cost is
\be
E_N^{(p)} =  \sum_{i=1}^N | r_i - b_i |^p
\ee
By using~\reff{P_k} and the Selberg integral~\reff{S_n} 
\begin{equation}
\begin{split}
S_2\left(k, N-k+1, \frac{ p}{2}\right)  
& = \left(\prod_{i=1}^2 \int_0^1 \ dx_i \, x_i^{k-1} (1-x_i)^{N-k}\right) \left |x_2-x_1\right |^{p} \\
& = \frac{\Gamma(k) \Gamma(N-k+1)\Gamma\left(k+\frac{p}{2}\right)\Gamma\left(N-k+1+\frac{p}{2}\right)\Gamma(1+ p)}{\Gamma\left(N+1+\frac{p}{2}\right) \Gamma(N+1+p)\Gamma\left(1+\frac{p}{2}\right)}\, .
\end{split}
\end{equation}
we get that the average of the $k$-th contribution is given by
\begin{equation}
\begin{split}
\overline{| r_k - b_k |^p} =  & \int_0^1 \ dx\, \int_0^1 \ dy\, P_k(x) \, P_k(y)\, |y-x|^p \\
= &  \left(  \frac{\Gamma(N+1)}{\Gamma(k) \, \Gamma(N-k+1)} \right)^2 \, \left(\prod_{i=1}^2 \int_0^1 \ dx_i \, x_i^{k-1} (1-x_i)^{N-k}\right) \left |x_2-x_1\right |^{p} \\
= & \left(  \frac{\Gamma(N+1)}{\Gamma(k) \, \Gamma(N-k+1)} \right)^2 \, S_2\left(k, N-k+1, \frac{p}{2}\right)\\
= & \frac{\Gamma^2(N+1)\Gamma\left(k+\frac{p}{2}\right)\Gamma\left(N-k+1+\frac{p}{2}\right)\Gamma(1+p)}{\Gamma(k) \Gamma(N-k+1) \Gamma\left(N+1+\frac{p}{2}\right) \Gamma(N+1+p)\Gamma\left(1+\frac{p}{2}\right) } \, \label{ass}
\end{split}
\end{equation}
and therefore we get the exact result
\begin{equation}
\begin{split}
\overline{E_N^{(p)}} & =  \frac{\Gamma^2(N+1)\Gamma(1+p)}{\Gamma\left(N+1+\frac{p}{2}\right) \Gamma(N+1+p)\Gamma\left(1+\frac{p}{2}\right) }   \sum_{k=1}^N\frac{\Gamma\left(k+\frac{p}{2}\right)\Gamma\left(N-k+1+\frac{p}{2}\right)}{\Gamma(k) \Gamma(N-k+1)} \\
& =  \frac{\Gamma\left(1 + \frac{p}{2}\right)}{p+1} \, N\, \, \frac{  \Gamma(N+1)}{\Gamma\left(N+1 + \frac{p}{2}\right)}
\end{split}
\end{equation}
where we made repeated use of the duplication and Euler's inversion formula for $\Gamma$-functions
\begin{subequations}
\begin{align}
\Gamma(z) \Gamma\left( z + \frac{1}{2} \right) & = 2^{1 - 2 z} \sqrt{\pi} \, \Gamma(2z) \\
\Gamma(1-z) \Gamma(z) & = \frac{\pi}{\sin( \pi z)}\, .
\end{align}
\end{subequations}
The exact result~\reff{ass} was known only in the cases $p=2, 4$ where the computation can be carried out simply by using the Euler Beta function~\reff{beta}, see~\cite{Caracciolo:169}.
With $p \neq 2, 4$, the average optimal cost has been computed only in the limit of large $N$~\cite{Caracciolo:169} at the order $o(N^{-1})$. From~\reff{ass} we easily get also the next correction in $1/N$
\be
\overline{E_N^{(p)}} = \frac{\Gamma\left(1 + \frac{p}{2}\right)}{p+1} \, N^{1-\frac{p}{2}} \, \left( 1 - \frac{p (p+2)}{8\, N}  + \frac{p (p+2) (p+4) (3p +2) }{384 \, N^2} + o(N^{-2}) \right)\, .
\ee

\section{Average cost in the TSP} \label{3}

Again consider two sets of $N$ points chosen at random in the interval $[0,1]$, the red $\{r_i\}_{\in[N]}$, and the blue $\{b_i\}_{i\in[N]}$, points. In the Travelling Salesman Problem (TSP) we have to choose a closed path which visits only once all the $2N$ points, alternating red and blue points along the walk, that is, see~\cite{Caracciolo:171}, two permutations $\sigma$ and $\pi$ in the symmetric group of $N$ elements ${\mathcal S}_N$, such that the total cost
\be
E_N^{(p)}[\sigma] := \sum_{i=1}^{N} [w_{\sigma(i),\pi(i)}^{(p)} + w_{\sigma\circ\tau(i),\pi(i)}^{(p)} ]= \sum_{i=1}^{N} \left[ | r_{\sigma(i)} - b_{\pi(i)} |^p + | r_{\sigma\circ\tau(i)} - b_{\pi(i)} |^p \right]
\ee
is minimal. In the previous equation $\tau$ is the shifting permutation, that is  $\tau(i) = i+1$ for $i\in[N-1]$ and $\tau(N)=1$.

\noindent When $p>1$ the optimal solution is given by the permutations~\cite{Caracciolo:171}
\be
\tilde{\sigma}(i) = 
\begin{cases}
2i-1 & i \leq  (N+1)/2 \\
2N -2i +2 & i > (N+1)/2 \label{sigmatilde}
\end{cases}
\ee
and
\begin{equation}
\tilde{\pi}(i) = \tilde{\sigma}(N+1-i)  =
\begin{cases}
2i  & i < (N +1)/2 \\
2N - 2i +1 & i \geq  (N +1)/2 \label{pitilde}
\end{cases}
\end{equation}
once both sets of points have been ordered. It follows that, in these cases, the optimal cost is
\be
E_N^{(p)} =  | r_{1} - b_1 |^p + | r_{N} - b_N |^p+ \sum_{i=1}^{N-1} [ | r_{i+1} - b_i |^p + | r_{i} - b_{i+1} |^p ]
\ee
By using~\reff{P_k} and the generalized Selberg integral~\reff{B_{j,k}} 
\begin{equation}
\begin{split}
& B_2\left(1,1; k, N-k,\frac{p}{2} \right) = \\
& =  \int_{0}^{1} dx_1 \, \int_{0}^{1} dx_2 \, x_1^{k-1} \, x_2^k\, (1-x_1)^{N-k} (1-x_2)^{N-k-1} \left|x_1-x_2 \right|^p \\
& = \frac{\left(k+\frac{p}{2}\right)\left(N-k+\frac{p}{2}\right)}{(N+p)\left(N+\frac{p}{2}\right)} \,S_2\left(k,N-k,\frac{p}{2} \right) \\
& = \frac{\Gamma(k)\Gamma(N-k) \, \Gamma(p+1) \, \Gamma\left(k+\frac{p}{2}+1\right) \, \Gamma\left(N-k+\frac{p}{2}+1\right)}{\Gamma(N+p+1) \, \Gamma\left(N+\frac{p}{2}+1\right) \, \Gamma\left( 1+ \frac{p}{2}\right)} \
\end{split}
\end{equation}
we get
\begin{equation}
\begin{split}
\overline{\left| b_{k+1}-r_k \right|^p} & = \overline{\left| r_{k+1}-b_k \right|^p} = \int_0^1 \, dx\, \int_0^1 \, dy\, P_k(x) \, P_{k+1}(y) \, \left| x-y \right|^p  \\
& = \frac{\Gamma^2(N+1)}{\Gamma(k) \, \Gamma(N-k) \, \Gamma(k+1)\, \Gamma(N-k+1)} \times\\
& \hphantom{=} \times \int_{0}^{1} dx \, dy \, x^{k-1} \, y^k (1-x)^{N-k} (1-y)^{N-k-1} \left|x-y \right|^p \\
& =  \frac{\Gamma^2(N+1)}{\Gamma(k) \, \Gamma(N-k) \, \Gamma(k+1)\, \Gamma(N-k+1)}\,  B_2\left(1,1; k, N-k,\frac{p}{2} \right)\\
& = \frac{\Gamma^2(N+1) \, \Gamma(p+1) \, \Gamma\left(k+\frac{p}{2}+1\right) \, \Gamma\left(N-k+\frac{p}{2}+1\right)}{\Gamma(k+1) \, \Gamma(N-k+1) \, \Gamma(N+p+1) \, \Gamma\left(N+\frac{p}{2}+1\right) \, \Gamma\left( 1+ \frac{p}{2}\right)} \,.
\label{ass2}
\end{split}
\end{equation}
%
%
from which we obtain
\begin{equation}
	\sum_{k=1}^{N-1} \overline{\left| b_{k+1}-r_k \right|^p} = 2 \, \Gamma(N+1) \Gamma(1+p) \left( \frac{(N+p+1) \, \Gamma(\frac{p}{2})}{4 (p+1) \, \Gamma(p) \, \Gamma(N+1+\frac{p}{2})}-\frac{1}{\Gamma(N+1+p)} \right).
\end{equation}
In addition
\begin{equation}
\begin{split}
\overline{\left| r_1 -b_1 \right|^p} = \overline{\left| r_N -b_N \right|^p} & = N^2 \int_{0}^{1} dx \, dy \, (xy)^{N-1} \left| x-y \right|^p \\
& = N^2 S_2\left(N,1,\frac{p}{2}\right) = \frac{N \, \Gamma(N+1) \, \Gamma(p+1)}{\left(N+\frac{p}{2}\right) \, \Gamma(N+p+1)} \,.
\end{split}
\end{equation}
Finally, the average optimal cost for every $N$ and every $p>1$ is
\begin{equation}\label{costtsp}
\begin{split}
\overline{E_N^{(p)}} &  = 2 \left(\overline{\left| r_1 -b_1 \right|^p} +\sum_{k=1}^{N-1} \overline{\left| b_{k+1}-r_k \right|^p} \right)\\
& = 2 \, \Gamma(N+1) \left[ \frac{(N+p+1) \, \Gamma \left( 1+ \frac{p}{2} \right)}{(p+1) \, \Gamma\left( N+1+\frac{p}{2} \right)} - \frac{2 \, \Gamma (p+1)}{(2N+p) \, \Gamma(N+p)} \right] \,.
\end{split}
\end{equation}
For $p=2$ this reduces to 
\begin{equation}
\overline{E_N^{(2)}} =  \frac{2}{3} \frac{N^2+4N-3}{(N+1)^2} \,.
\end{equation}
which was already known~\cite{Caracciolo:171}. For generic $p>1$ only the asymptotic behaviour for large $N$ was obtained in~\cite{Caracciolo:171} 
\begin{equation}
\lim\limits_{N \to \infty} N^{p/2-1} \overline{E_N^{(p)}} = 2\, \frac{\Gamma\left(\frac{p}{2}+1\right)}{p+1} \,,
\end{equation}
which agrees with the large $N$ limit of Eq. \eqref{costtsp}.

\begin{figure}[t]
	\centering
	\includegraphics[scale=1,
	keepaspectratio]{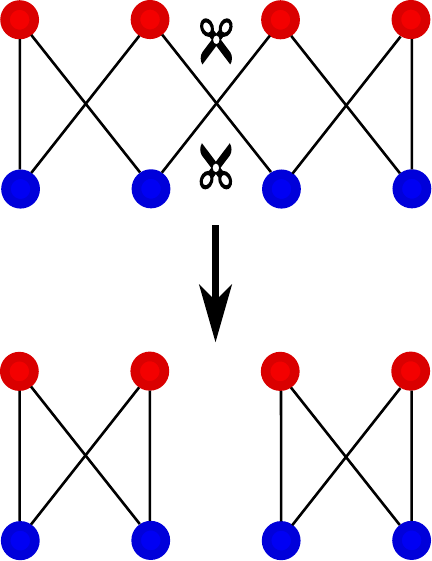} 
	\caption{Graphical representation of the cutting operation which brings from the optimal TSP cycle (top) to a possible optimal solution of the 2-factor problem (bottom). Here we have represented the $N=4$ case, where the cutting operation is unique. Notice that blue and red points are chosen on a interval, but here they are represented equispaced on two parallel lines to improve visualization.} \label{fig_cut2}
\end{figure}
\section{Cutting shoelaces: the 2-factor problem} \label{4}

In the 2-factor, or 2-assignment problem we carry out the minimization on the set of all the spanning subgraphs in which each vertex is shared by two edges. In different words, this problem corresponds to a loop covering of the graph, i.e. we relax the unique cycle condition we had in the TSP.
\begin{figure}[t]
\centering
\includegraphics[width=0.9\columnwidth]{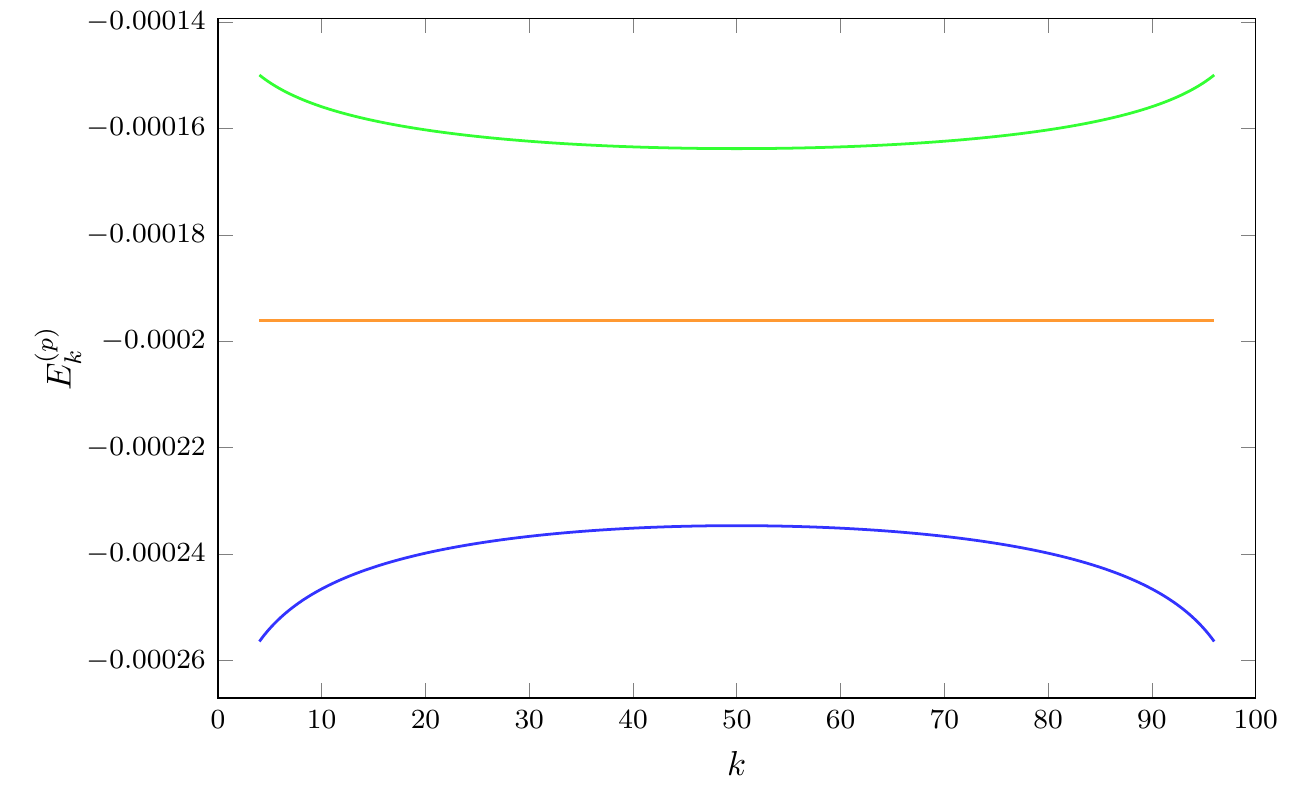} 
\caption{Plot of $E_k^{(p)}$ given in Eq.~\eqref{costcut} for various values of $p$: the green line is calculated with $p=2.1$, the orange with $p=2$ and the blue one with $p=1.9$; in all cases we take $N=100$.} \label{fig_cut}
\end{figure}
As shown in~\cite{Caracciolo:173}, for every value of $N$, the optimal 2-factor solution is always composed by an union of shoelaces loops with only two or three points of each color. As a consequence, differently from the assignment and the TSP cases, different instances of the disorder can have different spanning subgraphs that minimize the cost function. In particular these spanning subgraphs can always be obtained by ``cutting'' the optimal TSP cycle (see Fig. \ref{fig_cut2}) in a way which depends on the specific instance. 
This ``instance dependence'' makes the computation of the average optimal cost particularly difficult. However, one can show that the average optimal cost of the 2-factor problem is bounded from above by the TSP average optimal cost and from below by twice the assignment average optimal cost. Since in the large $N$ limit these two quantities coincide, one obtains immediately the large $N$ limit of the average optimal cost of the 2-factor problem.
Unfortunately, this approach is not useful for a finite-size system. But we can use Selberg integrals to obtain an upper bound: indeed we can compute the average cost obtained by ``cutting'' the TSP optimal cycle in specific ways.
When we cut at the $k$-position the optimal TSP into two different cycles we gain an average cost
\begin{equation}
E_k^{(p)} =   \overline{\left| b_{k+1}-r_k \right|^p} + \overline{\left| r_{k+1}-b_k \right|^p} - \overline{\left| b_{k+1}-r_{k+1} \right|^p} - \overline{\left| b_{k}-r_k \right|^p} \,.
\end{equation}
Once more, by using~\reff{P_k} and the generalized Selberg integral~\reff{B_{j,k}}, we obtain
\begin{equation}
\begin{split}
\lefteqn{ \overline{\left| b_{k}-r_k \right|^p} - \overline{\left| b_{k+1}-r_k \right|^p} = }\\
& = \frac{\Gamma^2(N+1) \, \Gamma(p+1) \, \Gamma \left( k+\frac{p}{2} \right) \, \Gamma \left( N-k+ \frac{p}{2} +1 \right)}{\Gamma(k) \, \Gamma(N-k+1) \, \Gamma(N+p+1) \, \Gamma\left( N + \frac{p}{2} +1 \right) \, \Gamma \left( \frac{p}{2} +1 \right)}  \left[ 1- \frac{k+ \frac{p}{2}}{k} \right] \\
& = - \frac{p}{2}\frac{\Gamma^2(N+1) \, \Gamma(p+1) \, \Gamma \left( k+\frac{p}{2} \right) \, \Gamma \left( N-k+ \frac{p}{2} +1 \right)}{\Gamma(k+1) \, \Gamma(N-k+1) \, \Gamma(N+p+1) \, \Gamma\left( N + \frac{p}{2} +1 \right) \, \Gamma \left( \frac{p}{2} +1 \right)} \,,
\end{split}
\end{equation}
and similarly
\begin{equation}
\begin{split}
\lefteqn{\overline{\left| b_{k+1}-r_{k+1} \right|^p} - \overline{\left| r_{k+1}-b_k \right|^p} = }\\ 
& = \frac{\Gamma^2(N+1) \, \Gamma(p+1) \, \Gamma \left( k+\frac{p}{2}+1 \right) \, \Gamma \left( N-k+ \frac{p}{2} \right)}{\Gamma(k+1) \, \Gamma(N-k) \, \Gamma(N+p+1) \, \Gamma\left( N + \frac{p}{2} +1 \right) \, \Gamma \left( \frac{p}{2} +1 \right)} \left[ 1- \frac{N-k+ \frac{p}{2}}{N-k} \right] \\
& = - \frac{p}{2}\frac{\Gamma^2(N+1) \, \Gamma(p+1) \, \Gamma \left( k+\frac{p}{2}+1 \right) \, \Gamma \left( N-k+ \frac{p}{2} \right)}{\Gamma(k+1) \, \Gamma(N-k+1) \, \Gamma(N+p+1) \, \Gamma\left( N + \frac{p}{2} +1 \right) \, \Gamma \left( \frac{p}{2} +1 \right)} \,.
\end{split}
\end{equation}
Their sum is
\begin{equation}\label{costcut}
E_k^{(p)}  = \frac{p}{2} \frac{\Gamma^2(N+1) \, \Gamma(p+1) \, \Gamma \left( k+\frac{p}{2} \right) \, \Gamma \left( N-k+ \frac{p}{2} \right)}{\Gamma(k+1) \, \Gamma(N-k+1) \, \Gamma(N+p) \, \Gamma\left( N + \frac{p}{2} +1 \right) \, \Gamma \left( \frac{p}{2} +1 \right)} \,,
\end{equation}
For $p=2$ this quantity is in agreement with what we got in~\cite{Caracciolo:173}
\begin{equation}
E_k^{(2)} =  \frac{2}{(N+1)^2}.
\end{equation}
For $p\neq2$, $E_k$ depends on $k$. In particular, for $1 < p < 2$ the cut near to 0 and 1 are (on average) more convenient than those near the center. For $p>2$ the reverse is true (see Fig.~\ref{fig_cut}).
Notice that for $p=2$ one can see that the best upper bound for the average optimal cost is given by summing the maximum number of cuts that can be done on the optimal TSP cycle. For $p\neq2$, however, this sum does not give a simple formula.

\section{Conclusions}
In this work we have been able to obtain some finite-size properties of a set of bipartite Euclidean optimization problems: the assignment, the bipartite TSP and the bipartite 2-factor problems by extensive use of the Selberg integrals. This confirms once more their importance and their wide application range. Interestingly, we have confirmed that the one dimensional 2-factor problem is more subtle to deal with than the other problems considered here, because even using Selberg integrals, we have not been able to find a finite-size upper-bound for the average optimal cost in the generic $p\neq2$ case. However, our approach allowed to understand the source of this difficulty: for $1 < p < 2$, the shortest loops tend to concentrate at the border of the $\left[ 0, 1\right]$ interval, while for $p > 2$ they tend to concentrate in the center of it.


\section*{References}

\bibliographystyle{unsrt}

\bibliography{Selberg}

\end{document}